\documentclass[pra
]{revtex4-2}
\usepackage{amsmath,amssymb,graphicx}
\usepackage{graphicx}

\begin{document}

\title{Rotating nonlinear states in trapped binary Bose-Einstein condensates
under the action of the spin-orbit coupling}
\author{Hidetsugu Sakaguchi$^{1}$ and Boris A. Malomed$^{2,3,}$\footnote{The corresponding author: malomed@tauex.tau.ac.il}}
\affiliation{$^1$Interdisciplinary Graduate School of Engineering Sciences, Kyushu
University, Kasuga, Fukuoka, 816-8580, Japan\\
$^{2}$Department of Physical Electronics, School of Electrical Engineering,
Faculty of Engineering, and Center for Light-Matter Interaction, Tel Aviv
University, Tel Aviv, 69978, Israel\\
$^{3}$Instituto de Alta Investigaci\'{o}n, Universidad de Tarapac\'{a}, Casilla 7D, Arica, Chile}

\begin{abstract}
We report results of systematic analysis of confined steadily rotating
patterns in the two-component BEC including the spin-orbit coupling (SOC) of
the Rashba type, which acts in the interplay with the attractive or
repulsive intra-component and inter-component nonlinear interactions and
confining potential. The analysis is based on the system of the
Gross-Pitaevskii equations (GPEs) written in the rotating coordinates. The
resulting GPE system includes effective Zeeman splitting. In the case of the
attractive nonlinearity, the analysis, performed by means of the
imaginary-time simulations, produces deformation of the known
two-dimensional SOC solitons (semi-vortices and mixed-modes). Essentially
novel findings are reported in the case of the repulsive nonlinearity. They
demonstrate patterns arranged as chains of unitary vortices which, at
smaller values of the rotation velocity $\Omega $, assume the straight
(single-string) form. At larger $\Omega $, the straight chains become
unstable, being spontaneously replaced by a trilete star-shaped array of
vortices. At still large values of $\Omega $, the trilete pattern rebuilds
itself into a star-shaped one formed of five and, then, seven strings. The
transitions between the different patterns are accounted for by comparison
of their energy. It is shown that the straight chains of vortices, which
form the star-shaped structures, are aligned with boundaries between domains
populated by plane waves with different wave vectors. A transition from an
axisymmetric higher-order (multiple) vortex state to the trilete pattern is
investigated too.
\end{abstract}

\maketitle

\section{Introduction}

Bose-Einstein condensates (BECs) in atomic gases offer a versatile platform
for the creation of various nonlinear modes in quantum matter. These
possibilities were realized experimentally for vortices \cite%
{JEWilliams,vort1,vort2,Kasamatsu}, effectively one-dimensional (1D)
matter-wave solitons \cite{Hulet,Salomon,Wieman}, including gap solitons
\cite{Oberthaler1}, weakly unstable (hence experimentally observable)
solitons of the Townes type in two dimensions (2D) \cite{Townes1,Townes2},
multidimensional \textquotedblleft quantum droplets" \cite%
{Tarruell1,Tarruell2,Inguscio,Salasnich} in binary BEC, whose stability is
provided by effects of quantum fluctuations \cite{Petrov,Petrov2}, etc.

The usual cubic (mean-field) self-attractive nonlinearity readily predicts
matter-wave solitons in the 2D and 3D free space, but they are made unstable
by the critical or supercritical collapse (in 2D or 3D, respectively) \cite%
{LBerge,Fibich,MalomedBA2}. Therefore, many works aimed to elaborate
conditions for the creation of \emph{stable} 2D and 3D matter-wave solitons.
An especially challenging problem is the stabilization of multidimensional
solitons with embedded vorticity, which, in the framework of basic settings,
are subject to a still stronger splitting instability \cite{MalomedBA2}.

A promising possibility for the creation of stable 2D \cite%
{Sakaguchi,Sherman,Umeda,ZhangY2,GautamS,ChenG,ZhongR,Sakaguchi2,Y.Li2,X.Chen,Deng}
and 3D \cite{HPu} solitons is offered by the binary BEC with the self- and
cross-attractive cubic interactions between its two components, which are,
in addition, linearly mixed, through their spatial gradients, by the
spin-orbit coupling (SOC). In the experiment, SOC is induced by a properly
designed set of laser beams illuminating the condensate \cite{SOC1,SOC2,SOC3}%
. The possibility to stabilize multidimensional matter-wave solitons by SOC
is upheld by the fact that SOC has been theoretically predicted \cite{half}
and experimentally realized \cite{2D SOC} in 2D, while the attractive sign
of the nonlinearity may be provided by the Feshbach resonance \cite%
{FR1,FR2,FR3,FR4}. Depending on the relative strength of the self- and
cross-component cubic attraction, the SOC systems give rise to absolutely
stable 2D \cite{Sakaguchi,Sherman} and metastable 3D \cite{HPu} solitons of
two types: semi-vortices (SVs) and mixed modes (MMs). Under the action of
SOC, stable SVs are built as compounds with zero vorticity in one component,
and vorticity $1$ in the other \cite{Sakaguchi,HPu}, or their mirror images,
with vorticities $-1$ and $0$ in the two components. The MM is produced,
approximately, as a combination of the SV and its mirror image.

In addition to the lowest-order (fundamental) SVs and MMs, the respective
system of coupled 2D Gross-Pitaevskii equations (GPEs) admits solutions for
\textit{excited states} of SVs and MMs, with equal vorticities added to both
components. In the case of the completely attractive nonlinear interactions,
all the excited states are unstable \cite{Sakaguchi}. They may be stabilized
in the system with opposite signs of the self-interaction in the two
components \cite{Deng}.

Another natural generalization of nonlinear SOC modes is to consider them in
the rotating form, with the centrifugal force balanced by an axisymmetric
trapping potential. This possibility was addressed in Refs \cite{rot1,rot2}
(without a relation to the SV and MM solitons) and \cite{Umeda} (the model
introduced in the latter work included the coupled GPEs in the form which
implied that the SOC-inducing set of laser beams was rotating, which is not
quite a realistic setup for the experiment). Here, we aim to obtain various
forms of stable rotating matter-wave modes, which are related, in
particular, to the SVs and MMs in the usual (non-rotating) situation
(assuming that the modes are set in rotation, while the underlying
SOC-inducing optical beams remain quiescent). To this end, the relevant
system of GPEs in the rotating reference frame in introduced in Section II.
The appropriate transformation casts the system in a form which is somewhat
similar to that derived in Ref. \cite{Umeda}, with an essential difference
that the transformed equations include effective Zeeman-splitting (ZS)
terms, which are induced by the rotation of BEC with respect to the
SOC-inducing optical framework. It is known that ZS may produce strong
effects in SOC systems \cite{ZS1,rot2,ZS2}.

Stable solutions, which are stationary in the rotating coordinates, are
obtained, for the attractive and repulsive signs of the nonlinear
interactions, in Sections III and IV, respectively. In the former case,
numerical analysis, performed by means of the imaginary-time simulations
\cite{imaginary,Bao}, produces, essentially, SV and MM solitons deformed by
the rotation. In the latter case (the repulsive interactions), the numerical
results demonstrate, first, formation of single- and multi-layer circular
vortex lattices, which resemble the patterns previously reported in Refs.
\cite{rot1,rot2,Umeda}. Most remarkable are completely novel patterns which
are revealed, in the case of the repulsive nonlinearity, by the systematic
numerical analysis, following the increase of the main control parameter,
\textit{viz}., the rotation angular velocity, $\Omega $. At first, it leads
to generation of multiple-vortex arrays which line up along a straight
string. Further increase of $\Omega $ destabilizes the rectilinear array of
vortices, replacing it by a star composed of three and, at still large
values of $\Omega $, of five and seven strings. The inferences produced by
this work are formulated in concluding Section V.

\section{The model}

The well-known 2D model of SOC of the Rashba type with strength $\lambda >0$%
\ is based on the following system of GPEs for wave functions $\phi _{\pm
}\left( X,Y,t\right) $ of the two components of the binary BEC, written in
the scaled form \cite{Sakaguchi},%
\begin{eqnarray}
i\frac{\partial \phi _{+}}{\partial t} &=&-\frac{1}{2}\nabla ^{2}\phi
_{+}+g\left( \left\vert \phi _{+}\right\vert ^{2}+\gamma \left\vert \phi
_{-}\right\vert ^{2}\right) \phi _{+}+U(r)\phi _{+}+\lambda \left( \frac{%
\partial }{\partial X}-i\frac{\partial }{\partial Y}\right) \phi _{-},
\notag \\
i\frac{\partial \phi _{-}}{\partial t} &=&-\frac{1}{2}\nabla ^{2}\phi
_{-}+g\left( \left\vert \phi _{-}\right\vert ^{2}+\gamma \left\vert \phi
_{+}\right\vert ^{2}\right) \phi _{-}+U(r)\phi _{-}+\lambda \left( -\frac{%
\partial }{\partial X}+i\frac{\partial }{\partial Y}\right) \phi _{+},
\label{GPE}
\end{eqnarray}%
where $\left( X,Y\right) $, with $\nabla ^{2}=\partial ^{2}/\partial
X^{2}+\partial ^{2}/\partial Y^{2}$, are the Cartesian coordinates in the
laboratory reference frame, $g$ is the overall strength of the nonlinear
interactions, $g<0$ and $g>0$ corresponding to the attraction and repulsion,
respectively, $\gamma \geq 0$ is the relative strength of the
inter-component interaction with respect to the self-interaction, and $U(r)$
is the trapping potential, which we adopt, as usual, in the
harmonic-oscillator form,%
\begin{equation}
U(r)=(1/8)(x^{2}+y^{2})  \label{U}
\end{equation}%
(the coefficient in this expression is fixed by means of scaling).

Coordinates $\left( x,y\right) $ in the reference frame rotating with
angular velocity $\Omega $ are defined as usual,%
\begin{eqnarray}
x &=&X\cos \left( \Omega t\right) +Y\sin \left( \Omega t\right) ,  \notag \\
y &=&Y\cos \left( \Omega T\right) -X\sin \left( \Omega t\right) ,  \label{XY}
\end{eqnarray}%
The GPE system (\ref{GPE}), transformed into the rotating coordinates (\ref%
{XY}), takes the following explicitly time-dependent form:
\begin{eqnarray}
i\frac{\partial \phi _{+}}{\partial t} &=&-\frac{1}{2}\nabla ^{2}\phi
_{+}+g\left( \left\vert \phi _{+}\right\vert ^{2}+\gamma \left\vert \phi
_{-}\right\vert ^{2}\right) \phi _{+}+U(\mathbf{r})\phi _{+}-\Omega \hat{L}%
_{z}\phi _{+}+\lambda e^{-i\Omega t}\left( \frac{\partial }{\partial x}-i%
\frac{\partial }{\partial y}\right) \phi _{-},  \notag \\
i\frac{\partial \phi _{-}}{\partial t} &=&-\frac{1}{2}\nabla ^{2}\phi
_{-}+g\left( \left\vert \phi _{-}\right\vert ^{2}+\gamma \left\vert \phi
_{+}\right\vert ^{2}\right) \phi _{-}+U(\mathbf{r})\phi _{-}-\Omega \hat{L}%
_{z}\phi _{-}+\lambda e^{i\Omega t}\left( -\frac{\partial }{\partial x}+i%
\frac{\partial }{\partial y}\right) \phi _{+},  \label{phi}
\end{eqnarray}%
where $\hat{L}_{z}=i\left( y\partial _{x}-x\partial _{y}\right) $ is the
angular-momentum operator, and the Laplacian is now written as $\nabla
^{2}=\partial ^{2}/\partial x^{2}+\partial ^{2}/\partial y^{2}$. The time
dependence in Eqs. (\ref{phi}) may be removed by the substitution%
\begin{equation}
\phi _{\pm }\equiv \psi _{\pm }\exp \left( \mp i\frac{\Omega }{2}t\right) ,
\label{phipsi}
\end{equation}%
which transforms the system into a form similar to the one which was
introduced in Ref. \cite{Umeda}, but with additional terms which are
tantamount to the effective ZS, with strength $\Omega /2$:%
\begin{eqnarray}
i\frac{\partial \psi _{+}}{\partial t} &=&-\frac{1}{2}\nabla _{+}^{2}\psi
+g\left( \left\vert \psi _{+}\right\vert ^{2}+\gamma \left\vert \psi
_{-}\right\vert ^{2}\right) \psi _{+}+U(\mathbf{r})\psi _{+}-\Omega \hat{L}%
_{z}\psi _{+}+\lambda \left( \frac{\partial }{\partial x}-i\frac{\partial }{%
\partial y}\right) \psi _{-}-\frac{\Omega }{2}\psi _{+},  \notag \\
i\frac{\partial \psi _{-}}{\partial t} &=&-\frac{1}{2}\nabla _{-}^{2}\psi
_{-}+g\left( \left\vert \psi _{-}\right\vert ^{2}+\gamma \left\vert \psi
_{+}\right\vert ^{2}\right) \psi _{-}+U(\mathbf{r})\psi _{-}-\Omega \hat{L}%
_{z}\psi _{-}+\lambda \left( -\frac{\partial }{\partial x}+i\frac{\partial }{%
\partial y}\right) \psi _{+}+\frac{\Omega }{2}\psi _{-}.  \label{psi}
\end{eqnarray}%
The ZS makes component $\psi _{+}$ energetically preferable in comparison
with $\psi _{-}$.

Various stationary states reported below were produced by imaginary-time
simulations of Eq. (\ref{psi}). The initial condition
\begin{equation}
\left( \psi _{+}\right) _{0}=0.5\exp (-x^{2}-y^{2}),~\left( \psi _{-}\right)
_{0}=0.3\exp (-x^{2}-y^{2})  \label{input}
\end{equation}%
was used for the simulations in most cases, unless indicated otherwise.

The system of coupled GPEs (\ref{psi}) has three usual dynamical invariants,
\textit{viz}., the total norm (proportional to the number of atoms in the
condensate),%
\begin{equation}
N=\int \int \left( \left\vert \psi _{+}\right\vert ^{2}+\left\vert \psi
_{-}\right\vert ^{2}\right) dxdy,  \label{N}
\end{equation}%
total angular momentum,%
\begin{equation}
M=\int \int \left( \psi _{+}^{\ast }\hat{L}_{z}\psi _{+}+\psi _{-}^{\ast }%
\hat{L}_{z}\psi _{-}\right) dxdy+\Omega N,  \label{M}
\end{equation}%
and energy%
\begin{gather}
E=\int \int \sum_{\pm }\left\{ \frac{1}{2}|\nabla \psi _{\pm }|^{2}+\frac{g}{%
2}|\psi _{\pm }|^{4}+U(r)|\psi _{\pm }|^{2}\right.  \notag \\
\lambda \left[ \psi _{+}^{\ast }\left( \frac{\partial \psi _{-}}{\partial x}%
-i\frac{\partial \psi _{-}}{\partial y}\right) +\psi _{-}^{\ast }\left( -%
\frac{\partial \psi _{+}}{\partial x}-i\frac{\partial \psi _{+}}{\partial y}%
\right) \right]  \notag \\
\left. -\Omega \left[ i\sum_{\pm }\psi _{\pm }^{\ast }\left( -x\frac{%
\partial }{\partial y}+y\frac{\partial }{\partial x}\right) \psi _{\pm }+%
\frac{|\psi _{+}|^{2}-|\psi _{-}|^{2}}{2}\right] \right\} dxdy,
\label{Energy}
\end{gather}%
where $\ast $ stands for the complex conjugate. Note that $M$ is conserved
in spite of the formally anisotropic form of the angular-momentum and SOC
operators in Eq. (\ref{psi}), cf. Refs. \cite{ZhongR,Deng}. Term $\Omega N$
in Eq. (\ref{M}) is the contribution to the angular momentum from the
overall rotation of the matter-wave state with respect to the laboratory
coordinates (in fact, this term is not essential, as $N$ is conserved
separately).

\section{Numerical results in the case of attractive interactions}

In this section, we consider the case of $g<0$, which corresponds to the
attractive nonlinearity in Eq. (\ref{psi}). The analysis starts from finding
linear eigenstates corresponding to $g=0$ and $\Omega >0$.

\begin{figure}[h]
\begin{center}
\includegraphics[height=4.cm]{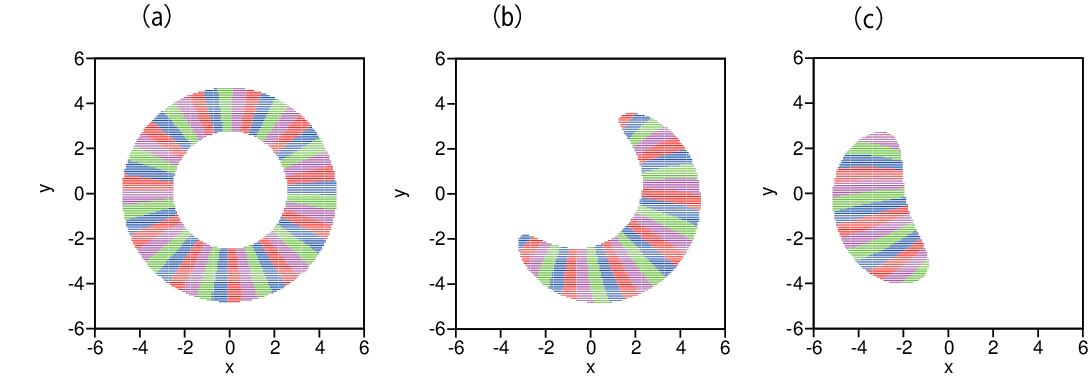}
\end{center}
\caption{(a) The numerically found vortex state for $g=-0.01$. (b) The
crescent-like state obtained for $g=-0.1$. (c) The semi-vortex soliton
obtained for $g=-0.5$. The other parameters are $\protect\gamma =0$, $\Omega
=0.3$, $\protect\lambda =2$, and $N=4$. Here and below (see Figs. \protect
\ref{fig4}-\protect\ref{fig8} and \protect\ref{fig15}-\protect\ref{fig13})
the area with $|\protect\psi _{+}|>0.08$ is plotted by different colors
indicating regions with Re$\protect\psi _{+}>0$ and Im$\protect\psi _{+}>0$
(green); Re$\protect\psi _{+}<0$ and Im$\protect\psi _{+}>0$ (blue); Re$%
\protect\psi _{+}<0$ and Im$\protect\psi +<0$ (red); Re$\protect\psi _{+}>0$
and Im$\protect\psi _{+}<0$ (magenta). This color coding makes it possible
to display both amplitude and phase profiles of $\protect\psi _{+}$ in the
single figure.}
\label{fig1}
\end{figure}

Figure \ref{fig1} shows stable vortex state produced by the numerical
solution for a set of increasing values of the self-attraction strength,
\textit{viz}., $g=-0.01$ (a), $g=-0.1$ (b), and $g=-0.5$ (c), fixing other
parameters and the norm as $\lambda =2$, $\gamma =0$, $\Omega =0.3$, and $N=4
$. The attractive interaction makes the wave function spontaneously
localized in the azimuthal direction, producing a crescent-like soliton in
Fig.~\ref{fig1}(b) and a still more compact one in Fig.~\ref{fig1}(c). The
transition from the nearly axisymmetric state to the one self-trapped in the
azimuthal direction is similar to that reported in Ref.~\cite{HS}

\begin{figure}[h]
\begin{center}
\includegraphics[height=4.cm]{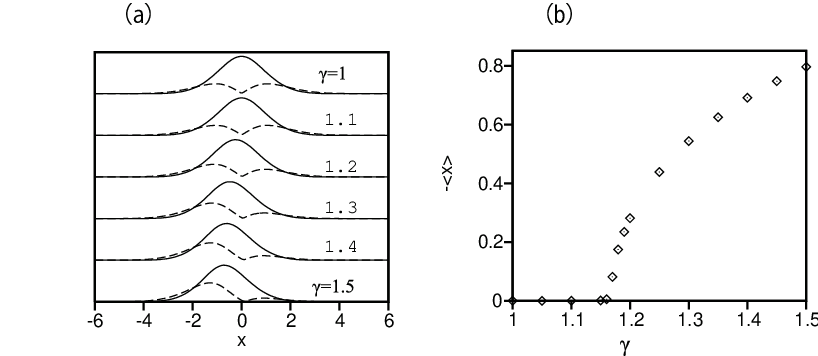}
\end{center}
\caption{(a) Profiles of $|\protect\psi _{+}(x,y)|$ and $|\protect\psi %
_{-}(x,y)|$ (solid and dashed lines) in the cross section $y=0$ for $\protect%
\gamma =1.5,1.4,1.3,1.2,1.1$, and $1$, fixing $g=-1$, $\Omega =0.35$, $%
\protect\lambda =0.5$, and $N=3$. (b) The average value $-\langle x\rangle $
(see Eq. (\protect\ref{average})) vs. $\protect\gamma $ for $g=1$, $\Omega
=0.35$, $\protect\lambda =0.5$, and $N=3$.}
\label{fig2}
\end{figure}

SV states in the attractive nonrotating system ($\Omega =0$) are stable at $%
\gamma <1$ (i.e., if the self-attraction is stronger than the
inter-component attraction) \cite{Sakaguchi}. These states persist in the
rotating system with $\Omega >0$. We study their evolution, following the
variation of parameter $\gamma $, as it is critically important for the SVs'
stability. Figure \ref{fig2}(a) shows profiles of $|\psi _{+}(x,y)|$ and $%
|\psi _{-}(x,y)|$ (solid and dashed lines, respectively) in the cross
section $y=0$ at $\gamma =1.5$, $1.4,$ $1.3,1.2,1.1$, and $1$, fixing $g=-1$%
, $\Omega =0.35$, $\lambda =0.5$, and $N=3$. Stable SVs are found at $\gamma
=1$ and $1.1$, satisfying their familiar property of vanishing the average
value of $x$ \cite{Sakaguchi} , i.e.,
\begin{equation}
\langle x\rangle =\int \int x|\psi _{+}|^{2}dxdy/\int \int |\psi
_{+}|^{2}dxdy=0.  \label{average}
\end{equation}%
In this vein, Fig. \ref{fig2}(b) shows $-\langle x\rangle $ as a function of
$\gamma $ for $g=-1$, $\Omega =0.35$, $\lambda =0.5$, and $N=3$. The SV
transition to an unstable states with $\langle x\rangle \neq 0$ occurs $%
\gamma =1.16$ (however, the instability does not transform the SVs into
MMs).

\begin{figure}[h]
\begin{center}
\includegraphics[height=3.5cm]{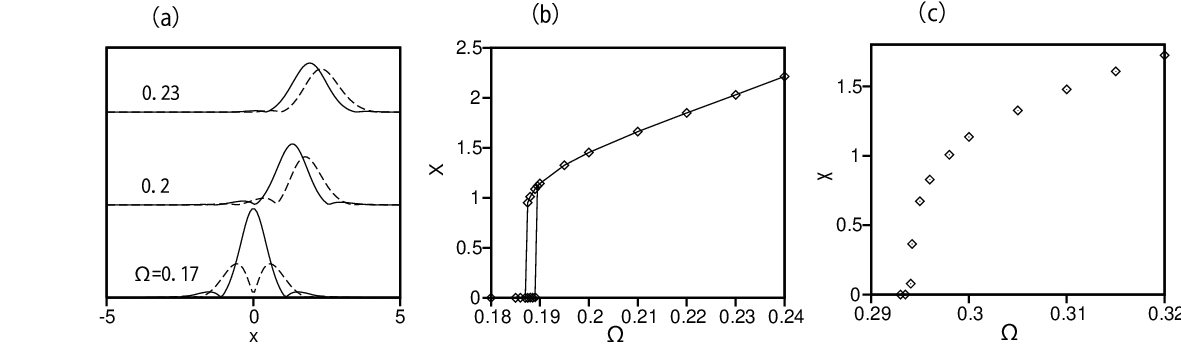}
\end{center}
\caption{(a) Profiles of $|\protect\psi _{+}(x)|$ and $|\protect\psi %
_{-}(x)| $ (solid and dashed lines) in the cross section $y=0$ at $\Omega
=0.17,0.2,$ and $0.23$ for $g=-1$, $\protect\gamma =0$, $\protect\lambda =2$%
, and $N=4$. (b) The $x$ coordinate of the center of mass, defined as per
Eq. (\protect\ref{X}) for $g=-1$, $\protect\gamma =0$, $\protect\lambda =2$,
and $N=4$. (c) The $x$ coordinate of the center of mass, defined as per Eq. (%
\protect\ref{X}) for $g=-1$, $\protect\gamma =0$, $\protect\lambda =1$, and $%
N=4$.}
\label{fig3}
\end{figure}

Further, Fig. \ref{fig3}(a) shows the profiles of $|\psi _{+}(x)|$ and $%
|\psi _{-}(x)|$ (the solid and dashed lines, respectively) in the cross
section $y=0$ for $\Omega =0.17,0.2$, and $0.23$, fixing
\begin{equation}
g=-1,\gamma =0,\lambda =2,N=4.  \label{parameters}
\end{equation}%
The usual stable SV state exists at $\Omega =0.17$, while the peak position
of $|\psi (x)|$ shifts to the right at $\Omega =0.2$ and $0.23$. For the
same parameter set (\ref{parameters}), Figure \ref{fig3}(b) shows the $x$
coordinate of the soliton's center of mass.
\begin{equation}
X=\int \int (|\psi _{+}|^{2}+|\psi _{-}|^{2})xdxdy/N,  \label{X}
\end{equation}%
with respect to the fact that $(|\psi _{+}|^{2}+|\psi _{-}|^{2})$ is the
total density of the condensate (cf. Eq. (\ref{N})). The center of mass
deviates from $x=0$ at $\Omega \geq 0.188$, featuring a weak hysteresis
revealed\ by the variation of $\Omega $. Figure \ref{fig3}(c) shows the $x$
coordinate of the soliton's center of mass for $g=-1$, $\gamma =0$, $\lambda
=1$, and $N=4$, which continuously deviates from $x=0$ in this case.

\begin{figure}[h]
\begin{center}
\includegraphics[height=3.5cm]{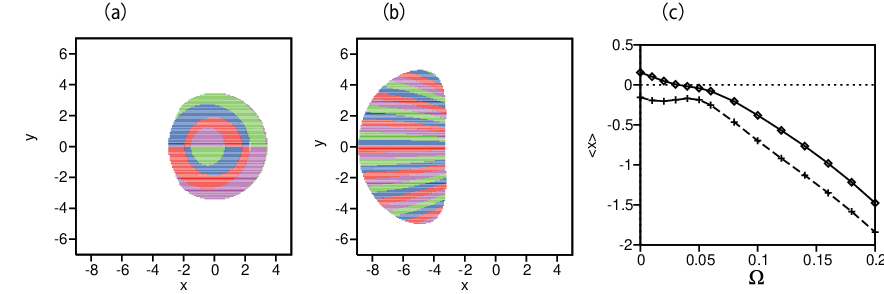}
\end{center}
\caption{Patterns of $|\protect\psi _{-}(x,y)|$ for the MM states at (a) $%
\Omega =0$ and (b) $\Omega =0.36$, for $g=-1$, $\protect\gamma =2$, $\protect%
\lambda =2$, and $N=1$. (c) Coordinates of the center of mass (\protect\ref%
{pm}) as a function of $\Omega $.}
\label{fig4}
\end{figure}

Figures \ref{fig4}(a) and (b) show wave patterns of the MM type of $\psi
_{-}\left( x,y\right) $ at (a) $\Omega =0$ and (b) $\Omega =0.36$ for $g=-1$%
, $\gamma =2$, $\lambda =2$, and $N=1$. Figure \ref{fig4}(a) produces a
stable MM state at $\Omega =0$. In this case, the center of mass of the MM
state is located at the origin. The MM state shown in Fig.~\ref{fig4}(a) for
$\Omega =0$ features a vortex centered at $x\simeq 0.66$ and $y=0$. As $%
\Omega $ increases, the center of mass shifts to the left, and soliton-like
state, similar to the one observed in Fig.~\ref{fig1}(c), appears in Fig.~%
\ref{fig4}(b) at $\Omega =0.36$. For large $\Omega $, the center of mass of
the localized state is driven off the origin by the centrifugal force.
Figure \ref{fig4}(c) shows the center-of-mass coordinates of the two
components,
\begin{equation}
\langle x_{\pm }\rangle =\frac{\int \int x|\psi _{\pm }|^{2}dxdy}{\int \int
|\psi _{\pm }|^{2}dxdy},  \label{pm}
\end{equation}%
(the solid and dashed lines; cf. Eq. (\ref{X})) as a function of $\Omega $.
At $\Omega =0$, $\langle x_{+}\rangle =0.156$ and $\langle x_{-}\rangle
=-0.156$, hence the center of the mass of the entire MM state is located at $%
x=0$, as said above. As $\Omega $ increases, $\langle x_{+}\rangle $
decreases monotonously, while $\langle x_{-}\rangle $ features a small peak
at $\Omega \simeq 0.05$ and then decreases.

\section{Numerical results in the case of repulsive interactions}

\begin{figure}[h]
\begin{center}
\includegraphics[height=3.5cm]{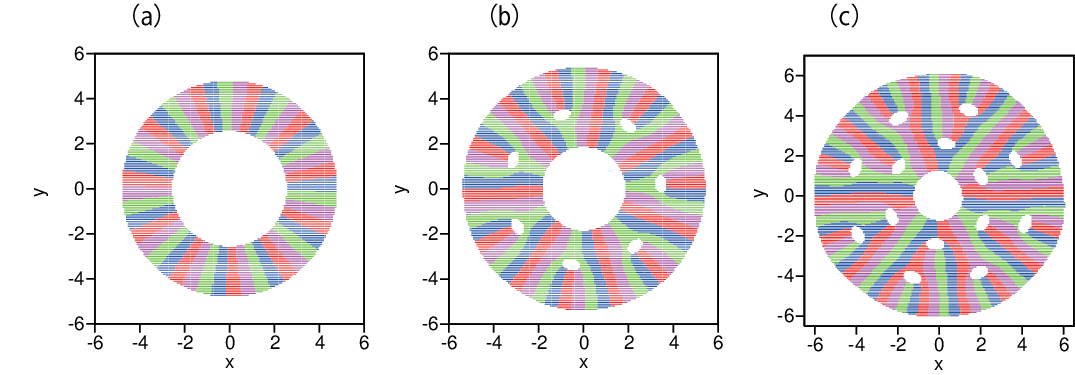}
\end{center}
\caption{Vortex patterns produced by the system with the repulsive
nonlinearity, \textit{viz}., $g=1$ (a), $g=10$ (b), and $g=30$ (c). The
patterns include higher-order vortices \ with vortex numbers (topological
charges) $\mathrm{VN}=11$, $7$, and $5$ located at the center, which are
surrounded by cricular layers built of unitary vortices (in panels (b) and
(c)). Other parameters and the norm are $\protect\gamma =0$, $\Omega =0.3$, $%
\protect\lambda =2$, and $N=4$.}
\label{fig5}
\end{figure}

In this section, we address the system with the repulsive nonlinearity, $g>0$%
. First, we address the transformation of the vortex patterns following the
variation of the nonlinearity strength, $g$.\ Figure \ref{fig5}(a) shows a
stable axisymmetric higher-order vortex state with the vortex number
(topological charge) $\mathrm{VN}=11$ for $g=1$, $\lambda =2$, $\gamma =0$, $%
\Omega =0.3$, and $N=4$. Figure \ref{fig5}(b) shows a circular vortex
lattice, built of $7$ unitary vortices, which surrounds the higher-order
vortex with $\mathrm{VN}=7$ located at the center, for $g=10$, $\gamma =0$, $%
\lambda =2$, $\Omega =0.3$, and $N=4$. Further, Fig. \ref{fig5}(c) shows a
more complex vortex lattice, including the outer and inner layers, built of $%
8$ and $6$ unitary vortices, respectively, and the higher-order vortex with $%
\mathrm{VN}=5$ located at the center, for $g=30$, $\gamma =0$, $\lambda =2$,
$\Omega =0.3$, and $N=4$. The repulsive interaction naturally drives
delocalization of the pattern and creation of additional vortices. We
categorize the vortex patterns in Figs.~\ref{fig5}(b) and (c) as two- and
three- layer structures, counting the central higher-order vortex. Actually,
the vortex layers appear at domain boundaries separating different areas of
the underlying background field. These vortex lattices are similar to ones
reported in previously studied models \cite{rot1,rot2,Umeda} (in particular,
the Ref. \cite{Umeda} dealt with a system similar to Eq. (\ref{psi}), but
without the effective ZS terms).

At a fixed value of $g$, the central vortex with $\mathrm{VN}>1$ appear
following the increase of $\Omega $. This is shown in detail in Fig. \ref%
{fig6}, for the gradual growth of $\Omega $: $0.18\rightarrow
0.19\rightarrow 0.20$ (note these these values are smaller than $\Omega
=0.30 $, for which Fig. \ref{fig5} is plotted). It is observed that two
unitary vortices, located near the center in panels (a) and (b), approach
each other and merge into the double vortex, with $\mathrm{VN}=2$, in panel
(c).
\begin{figure}[tbp]
\begin{center}
\includegraphics[height=3.5cm]{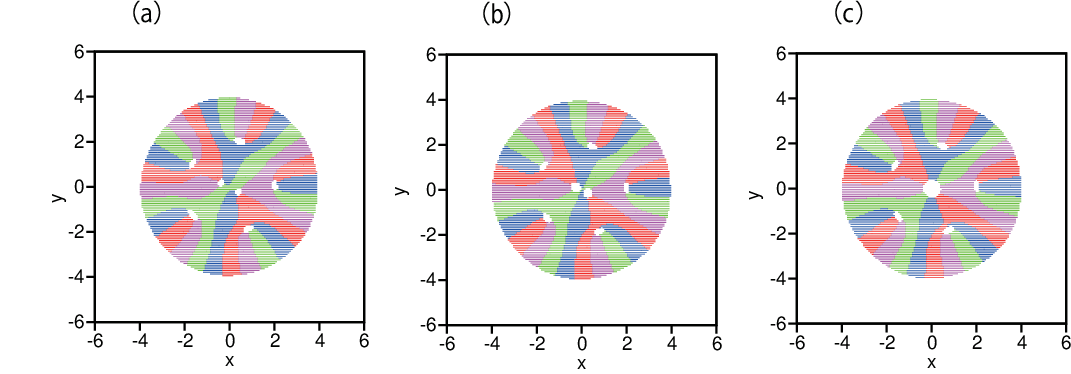}
\end{center}
\caption{Vortex patterns at (a) $\Omega =0.18$, (b) $0.19$, and (c) $0.2$
for $\protect\lambda =2$, $N=4$, $g=10$, and $\protect\gamma =0$. The
central vortex in (c) is a double one, with $\mathrm{VN}=2$.}
\label{fig6}
\end{figure}

\begin{figure}[h]
\begin{center}
\includegraphics[height=3.5cm]{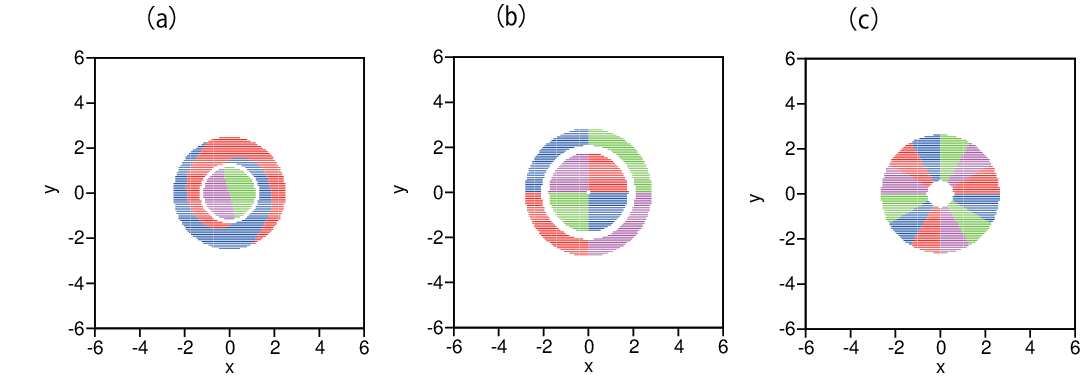}
\end{center}
\caption{(a) The $\protect\psi _{+}$ component of the SV state for $\Omega
=0.02$, $\protect\lambda =2$, $g=1$, $\protect\gamma =0$, and $N=4$. (b) The
same, but for $\Omega =0.08$ (four times as large as in (a)), $\protect%
\lambda =2$, $g=1$, $\protect\gamma =0$, and $N=4$. In this case, a vortex\
with $\mathrm{VN}=1$ is located at the center in the $\protect\psi _{+}$
component, and a vortex with $\mathrm{VN}=2$ is located at the center in $%
\protect\psi _{-}$ . (c) A higher-order vortex with $\mathrm{VN}=3$ at $%
\Omega =0.18$.}
\label{fig7}
\end{figure}
\begin{figure}[tbp]
\begin{center}
\includegraphics[height=3.2cm]{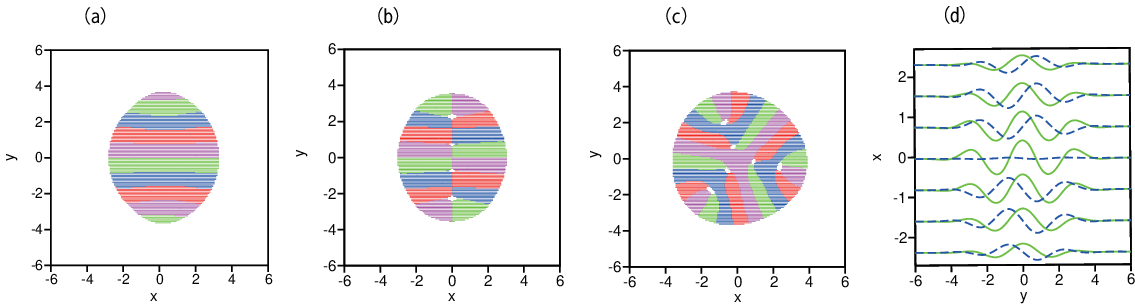}
\end{center}
\caption{Patterns of component $\protect\psi _{+}$ at (a) $\Omega =0$, (b) $%
\Omega =0.04$, and (c) $\Omega =0.16$ for $g=10$, $\protect\gamma =0$, $%
\protect\lambda =2$, and $N=4$. (d) Re$\protect\psi _{+}$ and Im$\protect%
\psi _{+}$ (solid green and dashed blue lines, respectively) at cross
sections drawn through $x=-2.49,-1.66,-0.83,0,0.83,1.66$, and $2.49$ for $%
\Omega =0.09$, in the case when the single array of unitary vortices is
aligned with the vertical axis, $x=0$.}
\label{fig8}
\end{figure}
\begin{figure}[tbp]
\begin{center}
\includegraphics[height=4.cm]{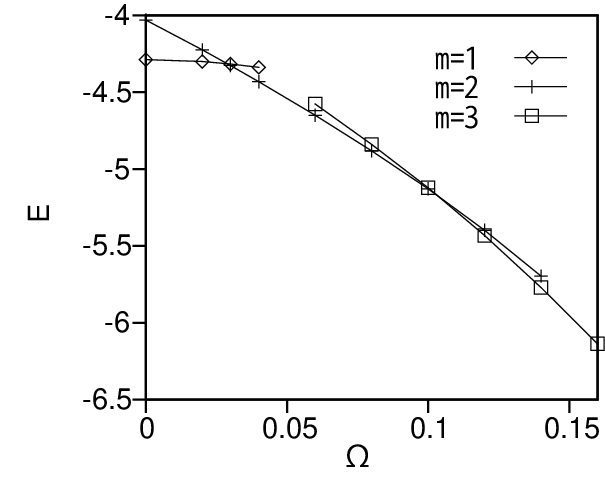}
\end{center}
\caption{The total energy $E$, defined as per Eq. (\protect\ref{Energy}), as
a function of the rotation angular velocity $\Omega $, for three types of
wave patterns, \textit{viz}., plane waves ($\mathrm{m=1}$), single-string
vortex arrays ($\mathrm{m=2}$), and trilete vortex arrays ($\mathrm{m=3}$),
for $\protect\lambda =2$, $g=10$, $\protect\gamma =0$, and $N=4$.}
\label{fig9}
\end{figure}
\begin{figure}[tbp]
\begin{center}
\includegraphics[height=4.cm]{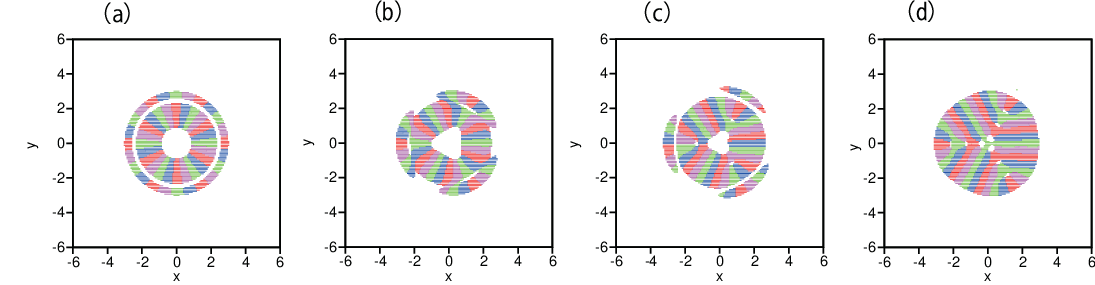}
\end{center}
\caption{The transition from the axisymmewtric higher-order vortex state to
the trilete pattern, following the variation of $g$, for $\protect\gamma =0$%
, $\protect\lambda =4$, $\Omega =0.1$, and $N=4$: (a) $g=0.5$, (b) $g=0.65$,
(c) $g=1$, and (d) $g=2$.}
\label{fig15}
\end{figure}
\begin{figure}[h]
\begin{center}
\includegraphics[height=3.2cm]{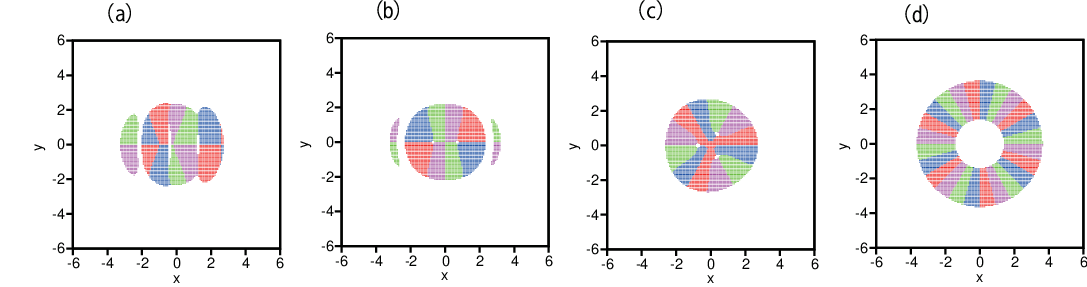}
\end{center}
\caption{The wave patterns in the $\protect\psi _{+}$ component at $\Omega
=0 $ (a), $\Omega =0.08$ (b) , $\Omega =0.16$ (c) , and $\Omega =0.24$ (d),
for $g=1$, $\protect\gamma =2$, $\protect\lambda =2$, and $N=4$.}
\label{fig10}
\end{figure}
\begin{figure}[h]
\begin{center}
\includegraphics[height=3.2cm]{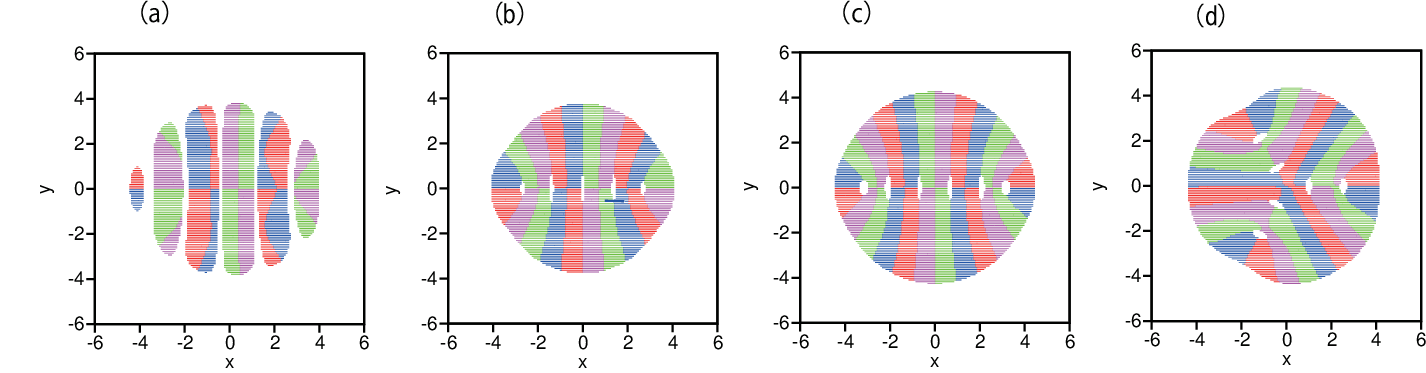}
\end{center}
\caption{(a) The standing-wave pattern of component $\protect\psi _{+}$ for $%
g=10$, $\protect\gamma =2$, $\Omega =0$, $\protect\lambda =2$, and $N=4$.
Wave patterns of $\protect\psi _{+}$ in the region of $|\protect\psi %
_{+}|>0.08$ are plotted by four colors, as defined in the caption to Fig.
\protect\ref{fig1}. (b) The wave pattern of $\protect\psi _{+}$ at $\Omega
=0.09$. (c) The wave pattern of $\protect\psi _{-}$ at $\Omega =0.09$. (d)
The wave pattern of $\protect\psi _{-}$ at $\Omega =0.1$. }
\label{fig11}
\end{figure}
\begin{figure}[tbp]
\begin{center}
\includegraphics[height=4.cm]{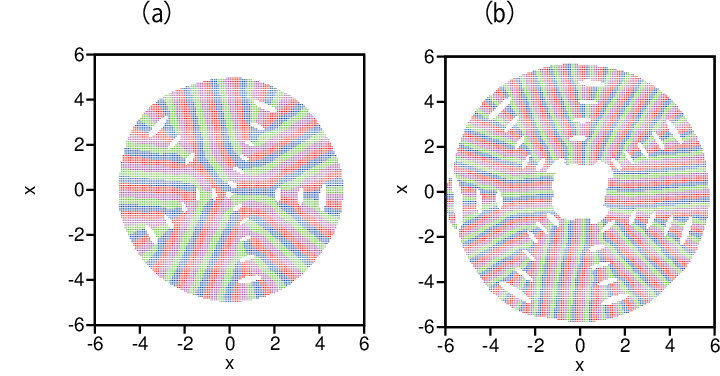}
\end{center}
\caption{The wave patterns at $\protect\lambda =5$ (a) and $\protect\lambda %
=8$ (b) for $g=10$, $\protect\gamma =2$, $\Omega =0.1$, and $N=4$. The
arrays of unitary vortices are attached to five and seven domain walls at $%
\protect\lambda =5$ and $\protect\lambda =8$, respectively.}
\label{fig13}
\end{figure}

Figure \ref{fig7} shows three patterns of $\psi _{+}\left( x,y\right) $ at
(a) $\Omega =0.02$, (b) $\Omega =0.08$, and (c) $\Omega =0.18$ for $\lambda
=2$, $g=1$, $\gamma =0$, and $N=4$. Figure \ref{fig7}(a) shows an SV state
with zero vorticity in the $\psi _{+}$ component, Fig.~7(b) shows a state
with vorticity $\mathrm{VN}=1$ in $\psi _{+}$, and in Fig.~7(c) one observes
a higher-order vortex with $\mathrm{VN}=3$ in $\psi _{+}$. In these cases,
the vorticity of $\psi _{-}$ is larger by $1$ than that of $\psi _{+}$,
which is the above-mentioned characteristic feature of SV states. For larger
$\Omega $, a central vortex appears with still larger values of $\mathrm{VN}$%
. Additional unitary vortices, which were observed above in Figs. \ref{fig5}%
(b) and (c), do not appear around the central multiple vortex in the present
case.

At large values of $g$, the flattening effect of the strong self-repulsion
gives rise to a nearly-plane-wave state at $\Omega =0$, in contrast to the
SV state observed at $g=1$. Actually, the approximate plane-wave state is
observed at $g\geq 1.5$, while the other parameters and the norm are fixed
as $\Omega =0.02$, $\lambda =2$, $\gamma =0$, and $N=4$. We address a
transition from the plane wave to structured states as $\Omega $ varies for
fixed $g=10$, $\lambda =2$, $\gamma =0$, and $N=4$. First, Fig. \ref{fig8}%
(a) exhibits the plane-wave state of component $\psi _{+}$ at $\Omega =0$,
there being no vorticity in this case. Next, Fig. \ref{fig8}(b) shows $\psi
_{+}$ at $\Omega =0.04$, exhibiting an array of unitary vorticities aligned
with the vertical axis, $x=0$. Next, Fig. \ref{fig8}(c) shows the structure
of $\psi _{+}$ at $\Omega =0.13$.

The vortex arrays, aligned with the single vertical straight line, $x=0$, in
Fig.~\ref{fig8}(b), or with the three lines, $x=0$ and $y=(\pm 1/\sqrt{3})x$
(at $x<0$) in Fig.~\ref{fig8}(c) (a trilete structure), are arranged by the
underlying background, which is composed of plane-wave domains with
different wave vectors. In particular, in Fig. \ref{fig8}(b) these are plane
waves approximated by $\exp \left( iky\right) $ at $x>0$ and $\exp \left(
-iky\right) $ at $x<0$. They are separated by the domain wall at $x=0$, to
which the vortex array is attached. The trilete state appears at $\Omega
\geq 0.12$. To illustrate the background structure, Fig. \ref{fig8}(d) shows
Re$\psi _{+}$ and Im$\psi _{+}$ (solid green and dashed blue lines,
respectively) at cross sections drawn through $%
x=-2.49,-1.66,-0.83,0,0.83,1.66$, and $2.49$, for the straight vortex
pattern at $\Omega =0.09$. The corresponding peak positions of Im$\psi _{+}$
are located to the right (left) of the peaks of Re$\psi _{+}$ for $y>0$ ($%
y<0 $). The wave pattern shown in Fig.~\ref{fig8}(c) is approximated by $%
\exp \{-i(\sqrt{3}k/2)x+i(k/2)y\}$ for $0<\theta <2\pi /3$, $\exp (-ik_{y})$
for $2\pi /3<\theta <4\pi /3$, and $\exp \{i(\sqrt{3}k/2)x+i(k/2)y\}$ for $%
4\pi /3<\theta <2\pi $, where $\theta $ is the angular coordinate in the $%
\left( x,y\right) $ plane. Accordingly, the three vortex arrays are aligned
with the domain boundaries between the background plane waves, corresponding
to $\theta =0$, $2\pi /3$, and $4\pi /3$. These vortex arrays are similar to
the ones reported in Refs. \cite{rot1,rot2}.

For the identification of the system's ground state in the case of
coexistence of different stationary modes, it is relevant to compute the
respective values of energy (\ref{Energy}) and thus find a state realizing
an energy minimum. Figure \ref{fig9} produces the results for three states:
the plane wave (denoted by $m=1$ in Fig. \ref{fig9}); the vortex array
aligned with the single boundary between two plane-wave background domains ($%
m=2$); and the trilete vortex array attached to boundaries between three
background domains ($m=3$). The energies of these three patterns are
exhibited in Fig. \ref{fig9} as functions of $\Omega $. These three species
of the patterns were produced by means of the imaginary-time simulations of
Eq. (\ref{psi}) with inputs taken, severally, as the plane wave at $\Omega
=0 $, the single-string vortex array at $\Omega =0.04$, and the trilete
array at $\Omega =0.16$. The total energy was computed if a given type of
the wave pattern was maintained in the course of the imaginary-time
simulations; the data were not recorded if the pattern spontaneously
switched in the course of the simulations. Thus, Fig. \ref{fig9}
demonstrates that the plane-wave state, denoted by the chain of rhombuses in
the figure, has the lowest energy at $\Omega <0.03$; the single-string
vortex array, which corresponds to the two-domain background, furnishes the
lowest energy at $0.03\leq \Omega \leq 0.1$; and the trilete vortex-array
pattern, based on the three-domain background, realizes the lowest energy
for $\Omega >0.1$. Thus, we conclude that the background with the increasing
number of domains becomes dominant as $\Omega $ increases.

The trilete pattern built of unitary vortices (similar to the one displayed
in Fig. \ref{fig8}(c)) can also develop from the original axisymmetric
higher-order vortex state as a result of the increase of the nonlinearity
coefficient $g$, for fixed values of $\gamma $, $\lambda $, $\Omega $, and $%
N $. Figure \ref{fig15} shows the corresponding patterns at $g=0.5$ (a), $%
g=0.65$ (b), $g=1$ (c), and $g=2$ (d) for $\gamma =0$, $\lambda =4$, $\Omega
=0.1$, and $N=4$. The stable axisymmetric higher-order vortex state with $%
\mathrm{VN}=6$ exists at $g=0.5$ and is maintained at $g=0.6$. At $g=0.65$,
three unitary vortices are generated at three points with some distance from
the origin, while the central vortex keeps $\mathrm{VN}=6$. The vortex
pattern gets deformed, reducing its rotational isotropy to the three-fold
symmetry. At $g=1$, three unitary vortices are released from the origin, and
a mixed state appears, including the central higher-order vortex state with
the remaining topological charge $\mathrm{VN}=3$ and the nascent trilete
structure. The distance of the satellite vortices from the origin increases
with the increase of $g$. At $g=2$, the central vortex disappears, and the
well-pronounced trilete state arises, as seen in Fig.~\ref{fig15}(d).

Next, we vary the rotation angular velocity $\Omega $ for fixed parameters $%
g=1$, $\gamma =2$, $\lambda =2$, and fixed norm $N=4$. Figure \ref{fig10}
displays the corresponding patterns of component $\psi _{+}$ at $\Omega =0$
(a), $\Omega =0.08$ (b), $\Omega =0.16$ (c), and (d) $\Omega =0.24$ for $g=1$%
. An MM state appears at $\Omega =0$. At $\Omega =0.08$ two and three
unitary vortices are located on the $x$ axis in the $\psi _{+}$ and $\psi
_{-}$ components, respectively, forming a single-string straight vortex
array. Three vortices are exhibited at $\Omega =0.16$ in Fig.~\ref{fig10}%
(c), with four vortices appearing in the respective pattern of component $%
\psi _{-}$ (one of them is located at the center). The latter pattern may be
categorized as the nascent trilete state, cf. Fig. \ref{fig15}(c). In Fig. %
\ref{fig10}(d), an axisymmetric higher-order vortex with $\mathrm{VN}=6$
exists at $\Omega =0.24$. Axisymmetric vortices with still larger values of $%
\mathrm{VN}$ appear for $\Omega \geq 0.24$, similar to the case of $\gamma
=0 $ shown in Fig.~\ref{fig7}.

The results for $g=10$ are displayed in Fig.~\ref{fig11}. A standing-wave
state appears at $\gamma =2$ and $g=10$, instead of the plane wave at $%
\gamma =0$ and $g=10$. Actually, this standing wave (at $g=10$) is similar
to the MM state at $g=1$. For larger $g$, the wave pattern naturally expands
due to the strong self-repulsion, remaining close to the standing wave.
Figure \ref{fig11}(a) shows the standing-wave state in component $\psi _{+}$
at $\Omega =0$. Figures \ref{fig11}(b) and (c) show wave patterns in the $%
\psi _{+}$ (b) and $\psi _{-}$ (c) components at $\Omega =0.09$. The nodal
line of the original standing wave is replaced by the vortex array, with the
total vorticity corresponding to $\mathrm{VN}=5$ for $\psi _{+}$ and $%
\mathrm{VN}=6$ for $\psi _{-}$. As above, $\mathrm{VN}$ of $\psi _{-}$ is
larger by $1$ than $\mathrm{VN}$ of $\phi _{+}$. Figure \ref{fig11}(d) shows
the pattern of $\psi _{-}$ at $\Omega =0.1$. The single-string vortex array
is unstable in this case, being replaced by the trilete pattern.

More complex vortex patterns than the trilete one are observed at larger
values of the SOC strength $\lambda $. Figure \ref{fig13} shows the wave
patterns for $\lambda =5$ (a) and $\lambda =8$ (b) with $g=10$, $\gamma =2$,
$\Omega =0.1$, and $N=4$. Vortex arrays are attached to five and seven
domain walls at $\lambda =5$ and $\lambda =8$, respectively.

\begin{figure}[tbp]
\begin{center}
\includegraphics[height=4.5cm]{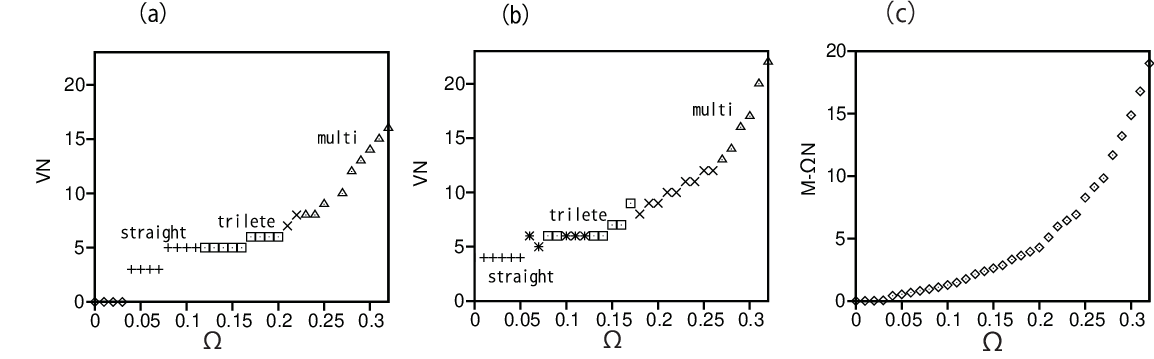}
\end{center}
\caption{(a) Vortex number $\mathrm{VN}$ as a function of rotation frequency
$\Omega $ for $\protect\gamma =0$, $\protect\lambda =2$, $g=10$, and $N=4$.
(b) $\mathrm{VN}$ vs. $\Omega $ for $\protect\gamma =2$, $\protect\lambda =2$%
, $g=10$, and $N=4$. (c) The reduced angular momentum $M-\Omega N$, defined
as per Eq. (\protect\ref{M}), vs. $\Omega $ for the same fixed parrameters
as in panel (a). The results summarized in this figure were produced by
means of the imaginary-time evolution method, with the input given by Eq. (%
\protect\ref{input}).}
\label{fig14}
\end{figure}

A general conclusion produced by the current analysis is that the pattern's
vortex number (topological charge), $\mathrm{VN}$, increases with the
increase of $\Omega $. The conclusion is summarized in Fig. \ref{fig14},
which shows the relationship between $\Omega $ and $\mathrm{VN}$ at $\gamma
=0$ (a) and $\gamma =2$ (b) for $\lambda =2$, $g=10$ and $N=4$. The data
plotted in Fig. \ref{fig14} were collected by the imaginary-time simulations
initiated with input given by Eq. (\ref{input}). The corresponding integer
value of $\mathrm{VN}$ was calculated as the sum of topological charges ($%
\mathrm{VN}$s) of the central axisymmetric vortex and unitary vortices
forming the arrays. As $\Omega $ increases, the plane-wave state takes place
at $\Omega \leq 0.04$ if $\gamma =0$. With the increase of $\Omega $, the
plane wave changes into the straight single-string vortex array, then the
trilete one, and, eventually, multi-string arrays. Marks \textquotedblleft
\textrm{x}" denote intermediate patterns including many unitary vortices.
For $\gamma =2$, the standing wave exists at $\Omega =0$. It changes into
the straight single-string array at small $\Omega $, then the trilete one,
and, eventually, multi-string patterns. Marks \textquotedblleft $\ast $"
denote a pattern formed by unitary vortices attached to four domain walls.

Finally, it is relevant to estimate characteristic parameters of the system
in physical units, assuming typical conditions which admit the experimental
realization of SOC \cite{SOC1,SOC2,SOC3}. For the 2D harmonic-oscillator
potential (see Eq. (\ref{U})) with trapping frequency $\sim 1$ Hz, SOC
spatial scale $\sim 10$ $\mathrm{\mu }$m, transverse trapping size $\sim 3$ $%
\mathrm{\mu }$m, and the number of atoms in the condensate $\sim 10^{4}$, we
conclude that the transitions from the straight (single-string) vortex
arrays to the trilete ones, and from the latter ones to the multi-string
arrays (see Fig. \ref{fig14}) may take place at the rotation angular
velocities $\Omega \sim 0.1$ Hz and $0.3$ Hz, respectively.

It is also natural to characterize the sequence of the vortex patterns by
the value of their angular momentum (\ref{M}), defined as per Eq. (\ref{M}).
Accordingly, Fig. \ref{fig14}(c) shows the reduced angular momentum, $%
M-\Omega N$, as a function of the rotation velocity $\Omega $ at $\gamma =0$%
, $\lambda =2$, $g=10$, and $N=4$, which corresponds to Fig. \ref{fig14}(a).
Naturally, the angular momentum is a monotonously growing function of $%
\Omega $.

\section{Conclusion}

The objective of this work is to develop the systematic numerical analysis
of steadily rotating states in the binary BEC under the action of SOC
(spin-orbit coupling) in the combination with the attractive or repulsive
intrinsic nonlinearity and axisymmetric trapping potential. The nonlinearity
includes cross- and self-interaction terms, with relative strength $\gamma $%
. To this end, the corresponding system of the GPEs (Gross-Pitaevskii
equations) is transformed into the rotating reference frame. The transformed
system includes effective ZS (Zeeman-splitting) terms. The systematic
numerical analysis, based on the imaginary-time simulations, demonstrates
deformation of the solitons of the SV (semi-vortex) and MM (mixed-mode)
types under the action of rotation, in the case of the attractive
nonlinearity. In the case of the repulsive interaction, various vortex
patterns appear. For the weakly repulsive interaction at $\gamma =0$, the SV
state exists at small values of the rotation velocity $\Omega $. It is
replaced by a vortex state at larger $\Omega $, which is further replaced by
a higher-order vortex state at even larger $\Omega $. For the strongly
repulsive interaction, a single-string array appears at relatively small $%
\Omega $, being replaced by a trilete array at larger $\Omega $. Eventually,
at still larger values of $\Omega $, stable star-shaped arrays emerge,
composed of five and seven strings. The transitions are explained by the
comparison of energy of the coexisting patterns of different types. The
rectilinear chains of vortices in single- and multi-string patterns are
attached to boundaries separating background domains filled by plane waves
with different wave vectors. The transition of the axisymmetric higher-order
vortex state to the trilete state was studied too, varying the nonlinearity
strength $g$ for fixed values of other parameters. In that case, unitary
vortices successively escape from the central higher-order one in three
directions, building the trilete pattern.

\section*{Acknowledgments}

The work of B.A.M. was supported, in a part, by the Israel Science
Foundation (grant No. 1695/22).


\begin{thebibliography}{99}
\bibitem{JEWilliams} J. E. Williams and M. J. Holl, Preparing topological
states of a Bose-Einstein condensate, Nature \textbf{401}, 568 (1999).

\bibitem{vort1} M. R. Matthews, B. P. Anderson, P. C. Haljan, D. S. Hall, C.
E. Wieman, and E. A. Cornell, Vortices in a Bose-Einstein condensate, Phys.
Rev. Lett. \textbf{83}, 2498-2501 (1999).

\bibitem{vort2} K. W. Madison, F. Chevy, W. Wohlleben, J. Dalibard, Vortex
formation in a stirred Bose-Einstein condensate, Phys. Rev. Lett. \textbf{84}%
, 806-809 (2000).

\bibitem{Kasamatsu} K. Kasamatsu, M. Tsubota, and M. Ueda, Vortex phase
diagram in rotating two-component Bose-Einstein condensates, Phys. Rev.
Lett. \textbf{91}, 150406 (2003).

\bibitem{Hulet} K. E. Strecker, G. B. Partridge, A. G. Truscott, and R. G.
Hulet, Formation and propagation of matter-wave soliton trains, Nature
\textbf{417}, 150-153 (2002).

\bibitem{Salomon} L. Khaykovich, F. Schreck, G. Ferrari, T. Bourdel, J.
Cubizolles, L. D. Carr, Y. Castin, and C. Salomon, Formation of a
matter-wave bright soliton", Science \textbf{296}, 1290-1293 (2002).

\bibitem{Wieman} S. L. Cornish, S. T. Thompson, and C. E. Wieman, Formation
of bright matter-wave solitons during the collapse of attractive
Bose-Einstein condensates, Phys. Rev. Lett. \textbf{96}, 170401 (2006).

\bibitem{Oberthaler1} B. Eiermann, Th. Anker, M. Albiez, M. Taglieber, P.
Treutlein, K.-P. Marzlin, and M. K. Oberthaler, Bright Bose-Einstein gap
solitons of atoms with repulsive interaction, Phys. Rev. Lett. \textbf{92},
230401 (2004).

\bibitem{Townes1} C.-A. Chen and C.-L. Hung, Observation of universal quench
dynamics and Townes soliton formation from modulational instability in
two-dimensional Bose gases, Phys. Rev. Lett. 125, 250401 (2020).

\bibitem{Townes2} B. Bakkali-Hassani, C. Maury, Y.-Q. Zhou, E. Le Cerf, R.
Saint-Jalm, P. C. M. Castilho, S. Nascimbene, J. Dalibard, and J. Beugnon,
Realization of a Townes soliton in a two-component planar Bose gas, Phys.
Rev. Lett. \textbf{127}, 023603 (2021).

\bibitem{Tarruell1} C. Cabrera, L. Tanzi, J. Sanz, B. Naylor, P. Thomas, P.
Cheiney, and L. Tarruell, Quantum liquid droplets in a mixture of
Bose-Einstein condensates, Science \textbf{359}, 301-304 (2018).

\bibitem{Tarruell2} P. Cheiney, C. R. Cabrera, J. Sanz, B. Naylor, L. Tanzi,
and L. Tarruell, Bright soliton to quantum droplet transition in a mixture
of Bose-Einstein condensates, Phys. Rev. Lett. 120, 135301 (2018).

\bibitem{Inguscio} G. Semeghini, G. Ferioli, L. Masi, C. Mazzinghi, L.
Wolswijk, F. Minardi, M. Modugno, G. Modugno, M. Inguscio, and M. Fattori,
Self-bound quantum droplets of atomic mixtures in free space?, Phys. Rev.
Lett. 120, 235301 (2018).

\bibitem{Salasnich} C. D'Errico, A. Burchianti, M. Prevedelli, L. Salasnich,
F. Ancilotto, M. Modugno, F. Minardi, and C. Fort, Observation of quantum
droplets in a heteronuclear bosonic mixture, Phys. Rev. Research \textbf{1},
033155 (2019).

\bibitem{Petrov} D. S. Petrov, Quantum mechanical stabilization of a
collapsing Bose-Bose mixture,. Phys. Rev. Lett. 115, 155302 (2015).

\bibitem{Petrov2} D. S. Petrov and G. E. Astrakharchik, Ultradilute
low-dimensional liquids, Phys. Rev. Lett. 117, 100401 (2016).

\bibitem{Townes} R. Y. Chiao, E. Garmire, and C. H. Townes, Self-trapping of
optical beams, Phys. Rev. Lett. 13, 479-482 (1964).

\bibitem{LBerge} L. Berg\'{e}, Wave collapse in physics: Principles and
applications to light and plasma waves, Phys. Rep. 303, 259 (1998).

\bibitem{Fibich} G. Fibich, The nonlinear Schr\"{o}dinger equation: singular
solutions and optical collapse, Berlin: Heidelberg; 2015.

\bibitem{MalomedBA2} B. A. Malomed, \textit{Multidimensional Solitons} (AIP
Publishing, Melville, NY, 2022).

\bibitem{Sakaguchi} H. Sakaguchi, B. Li, and B. A. Malomed, Creation of
two-dimensional composite solitons in spin-orbit-coupled self-attractive
Bose-Einstein condensates in free space, Phys. Rev. E 89, 032920 (2014).



\bibitem{Sherman} H. Sakaguchi, E. Ya. Sherman, and B. A. Malomed, Vortex
solitons in two-dimensional spin-orbit coupled Bose-Einstein condensates:
Effects of the Rashba-Dresselhaus coupling and the Zeeman splitting, Phys.
Rev. E \textbf{94}, 032202 (2016).

\bibitem{Umeda} H. Sakaguchi and K. Umeda, Solitons and Vortex Lattices in
the Gross--Pitaevskii Equation with Spin--Orbit Coupling under Rotation, J.
Phys. Soc. Jpn. \textbf{85}, 064402 (2016).

\bibitem{ZhangY2} Y. Zhang, M. E. Mossman, T. Busch, P. Engels, and C.
Zhang, Properties of spin-orbit-coupled Bose-Einstein condensates, Front.
Phys. \textbf{11}, 118103 (2016).

\bibitem{GautamS} S. Gautam and S. K. Adhikari, Vortex-bright solitons in a
spin-orbit-coupled spin-1 condensate, Phys. Rev. A \textbf{95}, 013608
(2017).

\bibitem{ChenG} G. Chen, Y. Liu, and H. Wang, Mixed-mode solitons in
quadrupolar BECs with spin-orbit coupling, Commun. Nonlinear Sci. Numer.
Simul. \textbf{48}, 318 (2017).

\bibitem{ZhongR} R. Zhong, Z. Chen, C. Huang, Z. Luo, H. Tan, B. A. Malomed,
Y. Li, Self-trapping under two-dimensional spin-orbit coupling and spatially
growing repulsive nonlinearity, Front. Phys. \textbf{13}, 130311 (2018).

\bibitem{Sakaguchi2} H. Sakaguchi and B. A, Malomed, One- and
two-dimensional gap solitons in spin-orbit-coupled systems with Zeeman
splitting, Phys. Rev. A \textbf{97}, 013607 (2018).

\bibitem{Y.Li2} Y. Li, X. Zhang, R. Zhong, Z. Luo, B. Liu, C. Huang, W.
Pang, and B. A. Malomed, Two-dimensional composite solitons in Bose-Einstein
condensates with spatially confined spin-orbit coupling, Commun. Nonlinear
Sci. Numer. Simul. \textbf{73}, 481 (2019).

\bibitem{X.Chen} X. Chen, Z. Deng, X. Xu, S. Li, Z. Fan, Z. Chen, B. Liu,
and Y. Li, Nonlinear modes in spatially confined spin-orbit-coupled
Bose-Einstein condensates with repulsive nonlinearity, Nonlinear Dynamics
101, 569 (2020).

\bibitem{Deng} H. Deng, J. Li, Z. Chen, Y. Liu, D. Liu, C. Jiang, C. Kong,
and B. A. Malomed, Semi-vortex solitons and their excited states in
spin-orbit-coupled binary bosonic condensates, Phys. Rev. E 109, 064201
(2024).

\bibitem{HPu} Y. Zhang, Z. Zhou, B. A. Malomed, and H. Pu, Stable solitons
in three dimensional free space without the ground state: Self-trapped
Bose-Einstein condensates with spin-orbit coupling, Phys. Rev. Lett. 115
253902 (2015).

\bibitem{SOC1} Y.-J. Lin, K. Jim\'{e}nez-Garc\'{\i}a, and I. B. Spielman,
Spin-orbit-coupled Bose-Einstein condensates, Nature 471, 83-86 (2011).

\bibitem{SOC2} B. M. Anderson, G. Juzeli\={u}nas, V. M. Galitski, and I. B.
Spielman, Synthetic 3D spin-orbit coupling, Phys. Rev. Lett. \textbf{108},
235301 (2012).

\bibitem{SOC3} V. Galitski and I. B. Spielman, Spin-orbit coupling in
quantum gases, Nature \textbf{494}, 49-54 (2013).

\bibitem{half} S. Sinha, R. Nath, and L. Santos, Trapped two-dimensional
condensates with synthetic spin-orbit coupling, Phys. Rev. Lett. \textbf{107}%
, 270401 (2011).

\bibitem{2D SOC} Z. Wu, L. Zhang, W. Sun, X.-T. Xu, B.-Z. Wang, S.-C. Ji, Y.
Deng, S. Chen, X.-J. Liu, and J.-W. Pan, Realization of two-dimensional
spin-orbit coupling for Bose-Einstein condensates, Science, 354, 83-88
(2016).

\bibitem{FR1} G. Roati, M. Zaccanti, C. D'Errico, J. Catani, M. Modugno, A.
Simoni, M. Inguscio, and G. Modugno, $^{39}$K Bose-Einstein condensate with
tunable interactions, Phys. Rev. Lett. \textbf{99}, 010403 (2007).

\bibitem{FR2} S. B. Papp, J. M. Pino and C. E. Wieman, Tunable miscibility
in a dual-species Bose-Einstein condensate, Phys. Rev. Lett. 101, 040402
(2008).

\bibitem{FR3} P. Zhang, P. Naidon and M. Ueda, Independent control of
scattering lengths in multicomponent quantum gases, Phys. Rev. Lett. 103
133202 (2009).

\bibitem{FR4} C. Chin, R. Grimm, P. Julienne, and E. Tiesinga, Feshbach
resonances in ultracold gases, Rev. Mod. Phys. 82, 1225-1286 (2010).

\bibitem{rot1} X.-Q. Xu and J. H. Han, Spin-orbit coupled Bose-Einstein
condensate under rotation, Phys. Rev. Lett. \textbf{107}, 200401 (2011).

\bibitem{rot2} X.-F. Zhou, J. Zhou, and C. Wu, Vortex structures of rotating
spin-orbit-coupled Bose-Einstein condensates, Phys. Rev. A \textbf{84},
063624 (2011).

\bibitem{ZS1} R. A. Duine and H. T. C. Stoof, Atom-molecule coherence in
Bose gases, Phys. Rep. \textbf{396}, 115-195 (2004).

\bibitem{ZS2} Y. V. Kartashov, V. V. Konotop, and F. K. Abdullaev, Gap
solitons in a spin-orbit-coupled Bose-Einstein condensate, Phys. Rev. Lett.
\textbf{111}, 060402 (2013).

\bibitem{imaginary} M. L. Chiofalo, S. Succi, and M. P. Tosi, Ground state
of trapped interacting Bose-Einstein condensates by an explicit
imaginary-time algorithm, Phys. Rev. E \textbf{62}, 7438-7444 (2000).

\bibitem{Bao} W. Z. Bao and Q. Du, Computing the ground state solution of
Bose-Einstein condensates by a normalized gradient flow, SIAM J. Sci. Comp.
\textbf{25}, 1674-1697 (2004).

\bibitem{MVakhitov} M. Vakhitov, and A. Kolokolov, Stationary solutions of
the wave equation in the medium with nonlinearity saturation, Radiophys.
Quantum Electron. 16, 783 (1973).

\bibitem{HS} H. Sakaguchi and B. A. Malomed. Solitons in combined linear and
nonlinear lattice potentials, Phys. Rev. A 81, 013624 (2010).

\bibitem{Noether} W. J. Thompson, \textit{Angular Momentum: An Illustrated
Guide to Rotational Symmetries for Physical Systems}, Volume 1 (Wiley-VCH,
Hoboken, New Jersey, 1994).
\end{thebibliography}
\end{document}